\documentclass[11pt]{article}
\pdfoutput=1
\usepackage{graphicx}
\usepackage{amssymb,amsmath}
\usepackage{bm} 
\usepackage{booktabs} 
\usepackage{array}
\usepackage{latexsym}
\usepackage{color}
\usepackage{slashed}
\usepackage{datetime}
\usepackage[nosort]{cite}
\usepackage{verbatim}
\usepackage{enumerate}
\usepackage{chngpage} 
\allowdisplaybreaks
\usepackage{psfrag}
\usepackage{mathtools}
\usepackage{mciteplus}
\usepackage{dsfont}

\usepackage[colorlinks=true,      linkcolor=blue,      urlcolor=blue,      
            filecolor=blue,      citecolor=blue,       pdfstartview=FitH,     
						pdfpagemode=UseNone,      bookmarksopen=true]{hyperref}  
\usepackage[all]{hypcap}     

\newcommand{\comm}[1]{} 

\def\({\left(}
\def\){\right)}
\def\[{\left[}
\def\]{\right]}

\def\pd{{\partial}}

\def\coeff#1#2{{\textstyle \frac{#1}{#2}}}

\def\One{{\hbox{ 1\kern-.8mm l}}}

\def\barray{\begin{array}}
\def\earray{\end{array}}
\def\be{\begin{equation}}
\def\ee{\end{equation}}
\def\bea{\begin{eqnarray}}
\def\eea{\end{eqnarray}}
\def\bal{\begin{align}}
\def\eal{\end{align}}

\numberwithin{equation}{section} 
\allowdisplaybreaks  

\makeatletter
\g@addto@macro\bfseries{\boldmath}
\makeatother

\definecolor{cardinal}{rgb}{0.6,0,0}
\definecolor{darkgreen}{rgb}{0,0.4,0}
\definecolor{golden}{rgb}{0.92, 0.7, 0}
\definecolor{midnight}{rgb}{0, 0, 0.5}
\definecolor{darkblue}{rgb}{0, 0, 0.7}
\definecolor{purple}{rgb}{0.5, 0, 0.5}

\def\cA{{\cal A}}

\def\cO{{\cal O}}

\def\tt{{\tilde{t}}}
\def\ty{{\tilde{y}}}
\def\tth{{\tilde{\theta}}}
\def\tpho{{\tilde{\varphi_1}}}
\def\tpht{{\tilde{\varphi_2}}}
\def\tph{{\tilde{\varphi}}}
\def\tx{{\tilde{x}}}

\usepackage{epsf}
\usepackage{cite}
\usepackage{fancyhdr}
\usepackage{bm}
\usepackage{enumerate} 

\usepackage{empheq} \empheqset{box=\fbox}
\usepackage[usenames,dvipsnames]{xcolor}

\numberwithin{equation}{section}  
\allowdisplaybreaks

\usepackage{graphicx}
\usepackage{tikz}
\usetikzlibrary{decorations.markings}
\tikzset{->-/.style={decoration={
			markings,
			mark=at position #1 with {\arrow{stealth}}},postaction={decorate}}}
\usepackage{adjustbox}
\usepackage{pgfplots}
\pgfplotsset{compat=1.11}
\usepgfplotslibrary{fillbetween}
\usetikzlibrary{intersections}

\pgfdeclarelayer{bg}
\pgfsetlayers{bg,main}

\begin{document}


\begin{flushright}

\end{flushright}

\vspace{3mm}

\begin{center}

  {\Huge {\bf Toroidal Tidal Effects in \\ \vspace{2mm} Microstate Geometries}}

\vspace{14mm}

{\large
\textsc{Nejc \v{C}eplak, Shaun Hampton, Yixuan Li}}
\vspace{12mm}

\textit{Institut de Physique Th\'eorique,
	Universit\'e Paris-Saclay,
	CNRS, CEA, \\ 	Orme des Merisiers, Gif-sur-Yvette, 91191 CEDEX, France  \\}  
\medskip

\vspace{4mm} 
%

{\footnotesize\upshape\ttfamily nejc.ceplak, shaun.hampton, yixuan.li @ ipht.fr} \\
\vspace{13mm}
 
\textsc{Abstract}

\end{center}

\begin{adjustwidth}{10mm}{10mm} 

\vspace{1mm}
\noindent

Tidal effects in capped geometries computed in previous literature  display no dynamics along internal (toroidal) directions.
However, the dual CFT picture suggests otherwise.
To resolve this tension, we consider a set of infalling null geodesics in a family of black hole microstate geometries with a smooth cap at the bottom of a long BTZ-like throat.
Using the Penrose limit, we show that a string following one of these geodesics feels tidal stresses along all  spatial directions, including  internal toroidal directions.
We find that the tidal effects along the internal directions are of the same order of magnitude as those along other, non-internal, directions.
Furthermore, these tidal effects oscillate as a function of the distance from the cap -- as a string falls down the throat it alternately experiences  compression and stretching. 
We explain some physical properties of this oscillation and comment on the dual  CFT interpretation.
\end{adjustwidth}

\thispagestyle{empty}
\clearpage



\baselineskip=14.5pt
\parskip=3pt

\tableofcontents

\baselineskip=15pt
\parskip=3pt

\section{Introduction}


The AdS/CFT duality \cite{Maldacena:1997re} is one of the major tools to study the black hole information paradox \cite{Hawking:1974sw}. 
The correspondence supports unitary black hole evolution and provides a dual description of black hole microstates within a non-gravitational theory.   
For example, a known family of three-charge horizonless geometries with a cap at the bottom of a long throat is dual to CFT states which, at the orbifold point, are constructed by acting with momentum-generating operators on Ramond-Ramond ground states obtained from length-one component strings \cite{Bena:2016ypk,Bena:2017xbt}.
%
%
Although such CFT states account only for a parametrically small fraction of the entropy \cite{Shigemori:2019orj,Mayerson:2020acj} of extremal black holes \cite{Strominger:1996sh}, they may offer insights about generic microstates of the corresponding black holes  within the bulk theory \cite{Lunin:2001jy,Mathur:2005zp,Skenderis:2008qn,Mathur:2009hf}.
Furthermore, away from the free orbifold point, the lowest possible mass gap in the strongly-coupled CFT matches that of energy excitations in such geometries  \cite{Bena:2018bbd}.%
\footnote{However, it is currently unclear how this small mass gap arises by deforming the CFT away from the free orbifold point.}

\begin{figure}[t]
	\label{fig:geodesic_in_geometry}
	\begin{adjustbox}{max totalsize={0.6\textwidth}{\textheight},center}
		\begin{tikzpicture}
			\begin{scope}[shift={(0,0)}]
				\draw[black,line width=1.5] (0, 1.5) arc (90:-90:1.5);
				\draw [ black, line width=1.5] (0, 1.5) --(-6,1.5);
				\draw [ black, line width=1.5] (0, -1.5) --(-6,-1.5);
				\draw [gray, thick, dashed] (0,1.5) -- (5, 1.5);
				\draw [gray, thick, dashed] (0,-1.5) -- (5, -1.5);
				\draw [gray, thick,dashed, ] (3,0) circle [x radius=0.3, y radius=1.5];
				\filldraw[fill=blue, opacity = 0.3, draw = none] (-1.5,1.5) -- (0,1.5) arc (90:-90: 0.3 and 1.5 ) -- (-1.5,-1.5) arc (-90:90: 0.3 and 1.5 ); 
				\draw [gray, thick,dashed] (1.5, -4) -- (1.5,0);
				\draw [gray, thick,dotted] (1.5, 5) -- (1.5,0);
				\draw [gray, thick,dashed] (-0.75, -4) -- (-0.75,-1.5);
				\draw [gray, thick,dashed] (-6, -4) -- (-6,-1.5);
				\draw [gray, thick,dotted] (0.0, 5) -- (0.0,1.5);
				\draw [gray, thick,dotted] (-1.5, 5) -- (-1.5,1.5);
				\draw [gray, thick,dotted] (-3.,3.5) -- (1.5,3.5);
				\node[] at (-0.75, -4.5) {\Large $r\sim \sqrt{n}\, a$};
				\node[] at (-6, -4.5) {\Large $r\sim \sqrt{n} \,b$};
				\node[] at (1.5, -4.5) {\Large $r = 0$};
				\draw [gray, thick,dashed](0.0,0) circle [x radius=0.3, y radius=1.5];
				\filldraw [gray, thick,dashed, fill= blue!15] (-1.5,0) circle [x radius=0.3, y radius=1.5];
				\draw[black, line width=1.5] (-6,-1.5) to[out=180, in=45] (-10, -4);
				\draw[black, line width=1.5] (-6,1.5) to[out=180, in=-45] (-10, 4);
				\draw [gray, thick,dashed](-10,0) circle [x radius=0.4, y radius=4];
				\node[align = center, centered] at (0.375, -3) {\Large Smooth\\\Large cap};
				\node[align = center, centered] at (-3.375, -3) {\Large AdS$_2\times S^1_y$ throat};
				\node[align = center, centered, fill=white] at (-8, -3) {\Large Asymptotic\\ \Large AdS$_3$ region };
				\node[align = center, centered] at (3.25, -3) {\Large Infinite BTZ\\ \Large throat};
				\draw[red, dashed, thick] (-10+0.4, 0) -- (-6,0) to[out = 0, in = 200] (-3.5,1.5) to[out = -20, in = 135]  (-1.8,-1.5) to[out=55, in=-110 ] (-0.8, 1.5) to[out=-70,in=100]  (0, -1.48) to[out=80, in=-95] (0.6, 1.41) to[out = -85, in = 93] (1, -1.1) to [out = 87, in= -92] (1.3, 0.75) to[out = -88, in = 91] (1.5,0); 
				\draw[red,  thick] (-10+0.4, 0) -- (-6,0) to[out = 0, in = 200] (-3.5,1.5);
				\draw[red,  thick]   (-1.8,-1.5) to[out=55, in=-110 ] (-0.8, 1.5);
				\draw[red,  thick]   (0, -1.5) to [out=80, in=-95] (0.6, 1.41);
				\draw[red,  thick]  (1, -1.1) to [out = 87, in= -92] (1.3, 0.75); 
			\end{scope}
			\begin{scope}[shift={(-0.7,3.5)}]
				\draw[scale=0.5,domain=-3.8:4.4, color=blue, smooth, variable=\x, thick]   plot ({\x},{3*exp(-\x*\x)*sin( 5* \x r)});
				\draw[scale=0.5,domain=-3:4.4, color=blue!40, dashed, smooth, variable=\x, thick]   plot ({\x},{3*exp(-\x*\x)});
				\draw[scale=0.5,domain=-3:4.4, color=blue!40, dashed, smooth, variable=\x, thick]   plot ({\x},{-3*exp(-\x*\x)});
			\end{scope}
		\end{tikzpicture}
	\end{adjustbox}
	\caption{
		A schematic depiction of the $(1,0,n)$ superstrata in the $r-y$ directions.
		The red curve depicts a trajectory of an infalling and spiralling null geodesic along which a string probe is travelling.
		The details of the microstructure are most prominent at the bottom of the throat (shaded in blue), where the tidal  forces are  greatest.
		The curve in blue above the geometry  illustrates the alternation between tidal stretching and compressing the string experiences as it approaches the cap region.}
\end{figure}
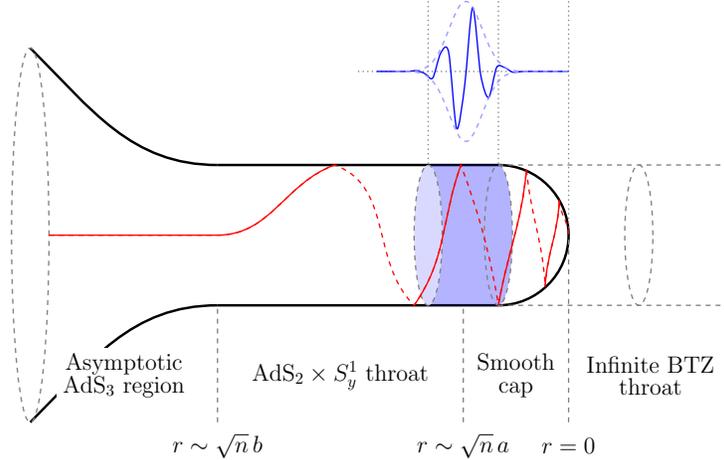

A well studied example is the extremal D1-D5-P black hole, with a known  match between the Bekenstein-Hawking entropy and the number of appropriate microscopic brane configurations  \cite{Strominger:1996sh}.
In the D1-D5 system, a large class of capped geometries that have a well-understood CFT description are  superstrata \cite{Bena:2015bea,Bena:2016ypk,Bena:2017xbt,Ceplak:2018pws, Heidmann:2019zws, Heidmann:2019xrd, Shigemori:2020yuo, Mayerson:2020tcl, Houppe:2020oqp},
which have the same charges and asymptotic structure as the black hole   \cite{Breckenridge:1996is}, but instead of having a horizon and an infinitely long throat, they cap off smoothly at the end of a long, but finite, \mbox{AdS$_2$ $\times S^1_y$} throat (see figure~\ref{fig:geodesic_in_geometry}). 
They are part of the microstate geometries programme which constructs smooth horizonless solutions within supergravity that represent  black hole microstates.


Non-trivial structure at the bottom of the throat implies physics that deviates from that of a classical black hole, ranging from tidal effects \cite{Tyukov:2017uig,Bena:2018mpb,Bena:2020iyw} to gravitational multipole moments \cite{Bena:2020see,Bianchi:2020bxa,Bena:2020uup,Bianchi:2020miz,Bah:2021jno}.
If capped geometries correspond to typical states of black holes, their inherent horizonless nature could lead to phenomena observable by future detectors \cite{Mayerson:2020tpn}, for example in gravitational echoes \cite{Dimitrov:2020txx}, as a massless probe falling into a horizonless geometry reaches the cap but is able to return to infinity \cite{Bena:2019azk, Bena:2020yii}.


%
Gravitational echoes are weakened if one considers a stringy probe.
Namely, tidal forces are able to excite  massive string modes, causing the probe to get trapped in the cap region \cite{Martinec:2020cml}.
In that paper, a massless string was sent along a null geodesic into an asymptotically \mbox{AdS$_3\times S^3\times T^4$} superstratum geometry.
It was found that the string gets tidally excited only along the $S^1_y$ direction of AdS$_3$ and the $S^3$, while there are no tidal forces acting along the toroidal directions.
This observation is in  tension with insights coming from the dual CFT.
There, excitations along the sphere can be described by fermionic degrees of freedom and excitations along the torus can be described by bosonic degrees of freedom \cite{Avery:2010qw}. 
Hence, if an infalling string gets tidally excited only along the $S^3$, the dual excitations are only fermionic in nature.
However, the initial probe propagating within the superstrata geometry has a precise CFT dual. 
Indeed, using deformations of the CFT, one can compute transitions from this state to bosonic and fermionic excitations. 
These transitions thus appear to be CFT signatures of tidal excitations of probes in the bulk \cite{Guo:2021ybz,Guo:2021gqd}. 
As a guiding principle, this suggests that a probe moving within the superstrata geometries should experience tidal excitations not only along the $S^3$, as shown in \cite{Martinec:2020cml}, but also along the $T^4$ directions, thus motivating the investigation of this paper.


We consider a null geodesic in the so-called $(1,0,n)$ superstratum \cite{Bena:2015bea, Bena:2016ypk, Bena:2017xbt}.
We choose the geodesic to be in the $\theta = \pi/2$ hypersurface of $S^3$.
Even though we also choose it to be directed radially in the asymptotic region, due to the momenta and angular momenta of the underlying geometry, the geodesic develops a spiral motion along  $S^1_y$ and  $S^3$ as it approaches the smooth cap (see figure~\ref{fig:geodesic_in_geometry}). 

The physical setup we have in mind is that of a massless string propagating along the chosen null geodesic.
To account for the interaction between the superstratum and the string probe, one needs to solve the string equations of motion.
However, to study the  tidal effects felt by a string as it moves along the geodesic, one can use the Penrose limit \cite{Blau:2002mw}
to capture local quadratic environment around a null geodesic.
After taking the limit, the metric can be written in the Brinkmann form and   the equations of motion for the $k$-th string oscillator mode in the light-cone gauge  can be written as 
\begin{align}
	\label{eq:StringEOM}
	\partial_{\tau}^2 w^i + {k^2} \,w^i + \left(\alpha'E\right)^2\cA_{ij}(\tau)\, w^j + i{k}\,{\alpha'E} \, B_{ij}(\tau)\,w^j = 0\,,
\end{align}
where $w^i$ denote directions orthogonal to the geodesic, $\tau$ is the world-sheet temporal coordinate and $E$ denotes energy.
The matrix $\cA_{ij}$  encodes tidal forces acting on the string.
In the above equation, it acts as a time dependent mass matrix, most importantly, if it is too negative, the system becomes unstable and massive string modes get excited.
In the above $B$ is the NS 2-form gauge potential, which will not be considered here.

Our results support the intuition from the CFT picture -- as a string propagates towards the cap, it encounters tidal stresses along all spatial directions, with  the  tidal forces in the toroidal directions being of the same order as those in the $S^1_y$ and $S^3$ directions. 
In addition, we show that the tidal effects display an oscillatory behaviour: a string alternately experiences compression and stretching. 
We analyse how these oscillations depend on the various parameters of the superstrata.
We also discuss the interpretation of tidal forces within the dual CFT.
There one can construct a simple state corresponding to a graviton moving within the superstrata geometry. 
By turning on an interaction of the theory one can then compute transitions of the initial probe into multiple excitations which carry polarizations along transverse directions such as the $S^3$ and the $T^4$.
These amplitudes exhibit a growth in time  which is suggestive of tidal effects from the perspective of the CFT.


In Section \ref{sec:MG}, we review the geometry of $(1,0,n)$ superstrata and analyse the structure of infalling null geodesics.
In Section \ref{sec:Tidal}, by using the Penrose limit around a null geodesic at $\theta = \pi/2$, we determine and analyse the tidal effects along the chosen trajectory.
In Section \ref{sec:CFT}, we discuss the implications of these tidal effects in the CFT picture. 
We finish with a conclusion and outlook on future work in Section~\ref{sec:CO}.

\section{Microstate geometry in the string frame}
\label{sec:MG}

In this section we describe a family of capped and horizonless  geometries, called $(1,0,n)$ superstrata which have the same charges as the \mbox{D1-D5-P} black hole.
Since our aim is to describe how a massless string 
interacts with the background as it falls towards the cap region, we consider the ten-dimensional metric in the string frame.
We analyse a particular set of  null geodesics and  rewrite the metric in a form that is useful in the study of tidal forces along  these geodesics.

\subsection{Background geometry}
\label{ssec:geo}

Consider the line element of  $(1,0,n)$ superstrata in the string frame \cite{Bena:2015bea, Bena:2016ypk,Bena:2017xbt}
\begin{align}
	\label{eq:SF10MetBump}
	ds_{10}^2 =  \Pi\left(\frac{1}{\Lambda}\, ds_6^2 ~+~ \sqrt{\frac{Q_1}{Q_5}}\, ds_{T^4}^2\right)\,,
\end{align} 
where $ds_6^2$ is the  six dimensional metric in the Einstein frame  and  $ds_{T^4}^2 = \delta_{ab}\,dz^a\,dz^b$ denotes the metric on the four-torus  $T^4$, which we take to be flat.
Typically the size of the latter   is taken to be much smaller than the size of the six-dimensional part of spacetime and thus  $T^4$ is often considered as an internal manifold \cite{Kanitscheider:2007wq}.
We focus on the near-horizon limit of the geometry in which $ds_6^2$ is asymptotically AdS$_3\times S^3$ 
\begin{align}
	ds_6^2 &=~ \sqrt{Q_1 Q_5} \, \Big( ds_{\rm AdS_3}^2 + ds_{S^3}^2 \Big)\,,
	\label{eq:EF6Dmet}
\end{align}
with $ds_{\rm AdS_3}^2$ and $ds_{S^3}^2$ denoting the asymptotically AdS$_3$ and  $S^3$ parts of the spacetime respectively.
For the $(1,0,n)$ superstrata these can be written as 
\begin{subequations}
	\label{eq:Met3Dab}
	\begin{align}
		ds_{\rm AdS_3}^2  &=~ \Lambda \, \bigg[ \frac{dr^2}{r^2 + a^2} +\frac{2\,r^2(r^2 + a^2)}{R_y^2 \, a^4}  \, dv^2 - \,\frac{1}{2\,  R_y^2\,A^4  G} \, \bigg( du + dv +  \frac{2   A^2 r^2}{a^2} \, dv\bigg)^{\!2}\,\bigg]  \,, 
		\label{eq:EF3dAdS}\\
		ds_{S^3}^2 &=~ \Lambda \, d\theta^2  +  \frac{\sin^2 \theta }{\Lambda}\,  \, \bigg(d \varphi_1  -  \frac{1}{\sqrt{2}\, R_y A^2}\,(du+dv) \bigg)^2 \nonumber\\*
		&\quad~  +  \frac{G\,\cos^2 \theta}{\Lambda}\,  \,  \bigg(d \varphi_2  +  \frac{1}{\sqrt{2}\, R_y\,a^2  A^2 \, G}\,\big(a^2(du-dv) - b^2 F \, dv \big) \bigg)^2\label{eq:EF3dS} \,,
	\end{align}
\end{subequations}
where we have used 
\begin{align}
	\label{eq:uvDef}
	u ~=~  \coeff{1}{\sqrt{2}} (t-y)\,, \qquad v ~=~  \coeff{1}{\sqrt{2}}(t+y) \,, 
\end{align}
with $t$ being the usual time coordinate and $y$ denoting the compact direction of AdS$_3$  with radius $R_y$ and is thus periodically identified as $y ~\sim ~  y ~+~ 2\pi  R_y $.

The detailed microscopic structure of the six dimensional geometry  is encoded in what we call bump functions
\begin{subequations}
	\label{eq:BumpDef1}
	\begin{align}
		F &~\equiv~  1 - \frac{r^{2n}}{(r^2+a^2)^{n}} \,, 
		&
		\Gamma& ~\equiv~ \sqrt{ 1 - \frac{b^2}{(2 a^2 +b^2)} \, \frac{r^{2n}}{(r^2 +a^2)^{n}}   }\,,\\
		G &~\equiv~ 1 - \frac{a^2 \, b^2}{2 a^2+ b^2} \, \frac{ r^{2n}}{(r^2+a^2)^{n+1}} \,,
		&
		\Lambda &~\equiv~ \sqrt{ 1 - \frac{a^2\,b^2}{(2 a^2 +b^2)} \, \frac{r^{2n}}{(r^2 +a^2)^{n+1}} \, \sin^2 \theta  } \,.
	\end{align}
\end{subequations}
However, the ten-dimensional metric in the string frame (\ref{eq:SF10MetBump}) involves  an additional bump function appearing as an overall conformal factor
\begin{align}
\label{eq:BumpDef2}
	\Pi \equiv \sqrt{ 1 + \frac{a^2\,b^2}{(2 a^2 +b^2)} \, \frac{r^{2n}}{(r^2 +a^2)^{n+1}} \, \sin^2 \theta  \,\cos{\left(\frac{2\sqrt{2} n v}{R_y} + 2\varphi_1\right)}}\,.
\end{align}
As we will show in the next sections, this function contains all information about tidal excitations in the internal directions.

Finally,  the geometry \eqref{eq:SF10MetBump}  contains  several parameters that characterize its properties.
The asymptotic radii of AdS$_3$ and $S^3$ are determined by $Q_1$ and $Q_5$, which denote the supergravity charges of the D1 and D5 branes in this system (see for example \cite{Skenderis:2008qn}).
Furthermore, the angular momenta and the momentum  are determined by the real-valued parameters $a$  and $b$ and an integer  $n$ as \cite{Bena:2017xbt}
\begin{align}
	\label{eq:ConsCharges}
	J_L  ~=~  J_R ~=~   \frac{R_y}{2} \,a^2\,,\qquad Q_P ~=~ \coeff{1}{2 } \, n \, b^2\,.
\end{align} 
However, not all parameters are independent as one has to impose 
\begin{align}
	Q_1 Q_5 ~=~  a^2 \, R_y^2\, A^2 \,, \qquad A ~\equiv~ \sqrt{ 1 +  \frac{b^2}{2 a^2}} \,,
	\label{SSreg1}
\end{align}
to ensure that the geometry is smooth everywhere \cite{Bena:2017xbt}.

Crucially,   $a$, $b$, and  $n$ determine the onset of the throat and cap regions in the geometry (see figure~\ref{fig:geodesic_in_geometry}) \cite{Bena:2016ypk, Martinec:2020cml}.
Focusing on the AdS$_3$ part of the metric, asymptotically  ($r \gg \sqrt{n}\, b$) the spacetime is approximately that of the extremal BTZ black hole and is thus locally AdS$_3$.
At $r \sim \sqrt{n}\,b$ the geometry transitions into a AdS$_2\times S^1_y$ throat, just like in the BTZ black hole. 
However, unlike in the case of the black hole, the throat region in the superstratum is  long, but finite and ends around $r \sim \sqrt{n}\, a$. 
The bottom of the throat is the region that contains most of the microstructure --  the bump functions have maxima/minima and consequently the superstratum significantly differs from an ordinary black hole. 
In fact, for the latter the throat is infinitely long, while for the superstratum the length of the throat is governed by the ratio $b/a$, which we usually take to be large in order to approximate the behaviour of black holes.
Finally, in the region $r \lesssim \sqrt{n}\, a$, \eqref{eq:EF3dAdS} smoothly caps off.


\subsection{Spiral infall along  null geodesics}

Our aim is to calculate the tidal forces felt by a massless string as it falls through the throat region towards the cap.
As such we are interested in null geodesics of \eqref{eq:SF10MetBump}, which are fully integrable \cite{Bena:2017upb}. 
For the metric in the string frame \eqref{eq:SF10MetBump} there are six Killing vector fields
 $K^{(2)}  \equiv (\pd/\pd \varphi_2)$, $ K^{(4)} \equiv  (\pd/\pd u)$ and $K^{(a)}= (\pd/\pd z^a)$, and two conformal Killing vector fields
$\tilde K^{(1)}  \equiv (\pd/\pd \varphi_1)$ and $\tilde K^{(3)} \equiv (\pd/\pd v)$, all of which can be used  to form eight scalars which are conserved along \emph{null} geodesics
\begin{align}
	L_1 \equiv 	\tilde K^{(1)}_M \frac{dx^M}{d\lambda}\,, &
	&	L_2 \equiv 	K^{(2)}_M 		\frac{dx^M}{d\lambda}\,, &
	&P \equiv \tilde	K^{(3)}_M \frac{dx^M}{d\lambda}\,,&
	&\hat E \equiv 	K^{(4)}_M \frac{dx^M}{d\lambda}\,,&
	&P_{a} \equiv  T^{(a)}_M \frac{dx^M}{d\lambda}\,,
		\label{eq:SFCC}
\end{align}
where $\lambda$ denotes the affine parameter along a null geodesic in the string frame. 
In addition, one can find a conformal Killing tensor \cite{Bena:2017upb} 
\begin{align}
	\label{eq:ckt}
	\Xi ~=~  
	Q_1 Q_5\, \Pi^2 \left(\frac{d\theta}{d\lambda}\right)^2 + \frac{L_1^2}{\sin^2\theta} + \frac{L_2^2}{\cos^2\theta}\,,
\end{align}
which can be shown to be conserved along null geodesics. Combining all these quantities with the null  condition
\begin{align}
	\label{eq:NullCond}
	g_{MN} \, \frac{dx^M}{d\lambda}   \frac{dx^N }{d\lambda}~=~  0\,,
\end{align}
allows us to fully parametrise null geodesics in $(1,0,n)$ superstrata.

In what follows we focus on the set of geodesics with 
\begin{align}
	\label{eq:ChaVal}
	L_1 ~=~ L_2 ~=~ 0\,, \qquad P ~=~ \hat E\,, \qquad P_a ~=~ 0\,, \qquad \Xi ~=~ 0\,.
\end{align}
Due to the highly non-trivial structure of the metric \eqref{eq:Met3Dab}, this choice does not mean that the motion is purely along the radial direction. 
In fact one can use \eqref{eq:SFCC}, \eqref{eq:ckt} and \eqref{eq:NullCond} to find
\begin{subequations}\label{eq:SFVels}
	\begin{gather}
		 \frac{du}{d\lambda} \,=\, -\frac{R_y\, A\,\left(a^2 + b^2 F\right)}{a\left(a^2 + r^2\right)  \Pi}\, \hat E   \,,\qquad 
		\frac{dv }{d\lambda} \,=\, - \frac{a\, R_y\, A}{\left(a^2 + r^2\right)  \Pi}\, \hat E \,,\qquad 
		\frac{dr }{d\lambda} \,=\, - \frac{ \sqrt{2}A \Gamma}{ \Pi} \, \hat E\,,\\
		\frac{d\theta}{d\lambda} ~=~ 0\,, \qquad 
		\frac{d\varphi_1}{d\lambda} ~=~ - \frac{\sqrt2\,a\, A \, \Gamma^2 }{ \left(a^2 + r^2\right) \Pi}\, \hat E \,,\qquad
		\frac{d\varphi_2}{d\lambda}~=~ 0 \,,\qquad 
		\frac{dz^a}{d\lambda} ~=~ 0\,,
	\end{gather}
\end{subequations}
which shows that even if the geodesic is directed radially in the asymptotic region of spacetime, as it traverses the throat, the momentum and angular momentum of the geometry cause the trajectory to rotate into other spatial directions  -- a massless particle travelling along such a geodesic towards the cap in a helical trajectory (see figure~\ref{fig:geodesic_in_geometry}).

We now perform a change of coordinates so that one of the coordinates is the affine parameter along the null geodesics with charges \eqref{eq:ChaVal}. We choose to replace the radial coordinate $r$ with $\lambda$, which induces
\begin{subequations}	\label{eq:CC1}
	\begin{gather}
		v  ~=~ v(\lambda) + \frac{\tt + \ty}{\sqrt2}\,, \qquad 
		u ~=~ u(\lambda) +  \frac{\tt - \ty}{\sqrt2}\,, \qquad 
		r ~=~ r(\lambda)\,,\\*
		\theta   ~=~ \frac{\pi}{2} - \tth\,, \qquad 
		\varphi_1 ~=~ \varphi_1(\lambda) + \tpho\,, \qquad 	
		\varphi_2 ~=~  \tpht\,, \qquad 
		z^a  ~=~ \tilde z^a\,, 
	\end{gather}
\end{subequations}
where the variables with a tilde correspond to independent coordinates labelling different null geodesics.%
\footnote{The choice to  shift in $\theta$ by $\pi/2$ is  for future convenience.} 
Inserting \eqref{eq:CC1} into \eqref{eq:SF10MetBump} and using \eqref{eq:SFVels} yields 
\begin{align}
	\label{eq:10dMetCC2}
	&ds^2_{10} ~ =~ d\lambda\,d\tt + \Pi~\Bigg\{\!\sqrt{Q_1 Q_5}\Bigg[\!
	- \frac{\left(a^2 + r^2\right) }{a^2 R_y^2 A^4 \, \Gamma^2}\, d\tt^2 
	+ \frac{\,r^2  \Gamma^2}{a^2 R_y^2  G}\left( d\ty + \frac{b^2  F}{2 a^2 A^2  \Gamma^2}d\tt\right)^2
	+  d\tth^2 \nonumber\\*
	&  
	+ \frac{\cos^2\tth}{\Lambda^2}\left(d\tpho - \frac{d\tt}{A^2 R_y}\right)^{\!2} \!
	+  \frac{G \sin^2\tth}{\Lambda^2}\left(d\tpht - \frac{ \Gamma^2}{R_y  G}\left(\frac{b^2  F}{2a^2 A^2  \Gamma^2}d\tt + d\ty\right)\right)^{\!\!2}\Bigg]\!+  \sqrt{\frac{Q_1}{Q_5}}\, ds_{T^4}^2\Bigg\},
\end{align}
where it should be understood that all bump functions are amended according to \eqref{eq:CC1} and furthermore, to avoid clutter we suppress the explicit dependence on  the affine parameter, thus all coordinates  without a tilde should be understood as being implicit functions $\lambda$ according to \eqref{eq:SFVels}. 
Note that we have also chosen $\hat E = \frac{1}{2\sqrt 2}$
to set the coefficient of $d \lambda \, d\tt$ to 1.
So far, this is just a rewriting of the initial metric  using a set of null geodesics, however this form is convenient when considering the Penrose limit to focus on the neighbourhood of a particular null geodesic, which in turn allows us to extract the tidal forces felt by a string as it moves along the chosen null trajectory.

\section{Tidal forces}
\label{sec:Tidal}

In this section we focus on a particular  null geodesic in the $(1,0,n)$ superstratum and use the Penrose limit to analyse its  neighbourhood.
Rewriting the resulting metric in the Brinkmann form allows us to extract information about the tidal forces felt by an object travelling along the chosen geodesic.
We find that there exist non-vanishing tidal forces along the $T^4$ directions, stemming from the overall conformal factor appearing in the string frame metric.

\subsection{Penrose limit}
\label{ssec:PL}

\begin{figure}[t]
	\centering
	\includegraphics[width=\textwidth]{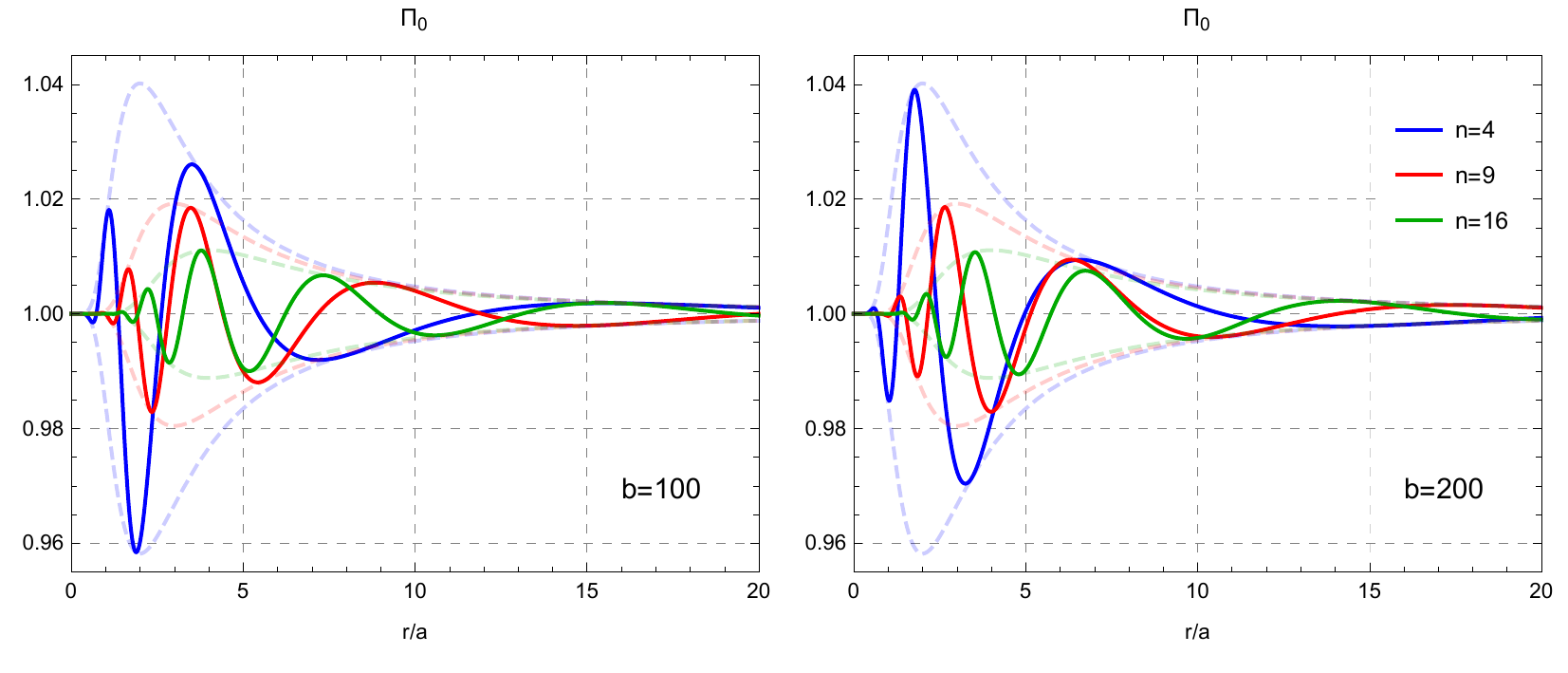}
	\caption{
		Plots of  $\Pi_0$ as a function of $r$. 
		Plots on the left correspond to $\tilde b \equiv b/\sqrt2 a =100$, while $\tilde b = 200$ on the right. 
		In each case we plot the curves for three different values of $n = 4,9,16$. 
		The dashed curves represent the envelopes of oscillations.}
	\label{fig:Pi0Plot}
\end{figure}

Let us now consider a null geodesic with \eqref{eq:ChaVal} at $\theta = \pi/2$ or more equivalently $\tth = 0$, and  $\ty = \tpho = \tpht = 0$. 
To extract the behaviour of \eqref{eq:10dMetCC2} in the neighbourhood of this geodesic, we use the Penrose limit \cite{Blau:2002mw} which amounts to  rescaling%
\footnote{Note that at $\tth = 0$, the $\tpht$ circle pinches off, hence  the latter  must not be rescaled in the Penrose limit as the relevant focusing is already contained within the scaling of $\tth$. 
}
\begin{align}
		\label{eq:PenLim}
		\lambda ~\to~ \lambda\,, \qquad
		\tt ~\to~ \Omega^2 \, \tt\,,\qquad 
		\left(\ty\,, \tpho\,, \tth\right) ~\to~ \Omega \left(\ty\,, \tpho\,, \tth\right)\,,\qquad 
		\tpht ~\to~ \tpht\,,
	\end{align}	
followed by taking $\Omega \to 0$ and retaining only the terms in the metric  which scale as $\Omega^2$.
Applying the above scaling to \eqref{eq:10dMetCC2} yields 
\begin{align}
	ds^2_{10} ~=~ d\lambda\, d\tt + \Pi_0 \sqrt{Q_1 Q_5}  \left(\frac{r^2 \Gamma^2}{a^2 R_y^2 G} d\ty^2+ \frac{1}{G} d\tpho^2+ d\tth^2 +\tth^2 d\tph_2^2    \right)
	+ \Pi_0\sqrt{\frac{Q_1}{Q_5}}\, \delta_{ab}\,d\tilde z^a d\tilde z^b,
	\label{eq:10MetPL}
\end{align}
where we defined 
\begin{align}
	\Pi_0 ~\equiv~ \sqrt{ 1 + \frac{a^2 b^2}{2a^2 + b^2} \, \frac{r^{2n}}{\left(a^2 + r^2\right)^{n+1}} \, \cos{\left(\frac{2\sqrt{2} n v}{R_y} + 2\varphi_1\right)}}\,,\label{eq:Pi0}
\end{align}
which is the leading behaviour of \eqref{eq:BumpDef2} in the Penrose limit ($\Pi \to \Pi_0 + \cO(\Omega)$).
It is important to recall that after the change of coordinates \eqref{eq:CC1} $r$, $v$, and $\varphi_1$ appearing in $\Pi_0$ (as well as other bump functions appearing in \eqref{eq:10MetPL}) are  all implicitly  functions of $\lambda$.
Nonetheless, using \eqref{eq:SFVels}, one can express $\Pi_0$ as a function of $r$ and we depict its behaviour for some values of $b$ and $n$ in figure~\ref{fig:Pi0Plot}.
By relabelling the eight coordinates transverse to $\tt$ and $\lambda$ as
$\tilde x^i  =(\ty, \tpho, \tth \cos\tph_2, \tth \sin\tph_2, \tilde z^1, \tilde z^2,\tilde z^3,\tilde z^4)$,
where in particular we introduced a set of polar coordinates  $\tx^3 \equiv \tth \cos\tph_2$ and $\tx^4 \equiv \tth \sin\tph_2$, then  we can rewrite \eqref{eq:10MetPL} in a diagonal form
\begin{align}
	ds^2_{10} &~\equiv~ d\lambda\, d\tt + \sum_{i,j=1}^8 a_i^2(\lambda)\, \delta_{ij}\, d\tx^i d\tx^j\,,
	\label{eq:10MetPLDiag}
\end{align}
where all diagonal entries  are functions of $\lambda$ only.%
\footnote{Albeit implicitly through $r$, $v$,  and $\varphi_1$.}
For the analysis of the tidal forces acting on a string, it is convenient to rewrite the metric in the Penrose limit in the Brinkmann form
 \cite{Blau:2002mw}%
\footnote{Our sign convention for the matrix $\cA_{ij}$ follows \cite{Martinec:2020cml}, which is opposite of what is used in  \cite{Blau:2002mw}.}
\begin{align}
	\label{eq:BrinkForm}
	ds_{10}^2 ~=~ 2 dx^+\, dx^- - \left(\sum_{i,j=1}^8 \,\cA_{ij}(x^-)\,w^i\, w^j\right)\, (dx^-)^2 +  \sum_{i,j=1}^8 \delta_{ij}\, dw^i dw^j\,.
\end{align}
Since \eqref{eq:10MetPLDiag} is diagonal, the change of coordinates that yields the desired form of the metric is 
\begin{align}
	\label{eq:Brinktrans}
	\lambda ~=~ 2 x^-\,, \qquad \tt ~=~ x^+ + \frac12\sum_{i,j=1}^8 \frac{a_i'(x^-)}{a_i(x^-)}\, \delta_{ij}\, w^i\, w^j\,, \qquad  \tilde x^i ~=~ \frac{w^i}{ a_i(x^-)}\,,
\end{align}
from which it follows that the matrix $\cA_{ij}$ is also diagonal 
\begin{align}
	\label{eq:cAmet}
	\cA_{ij}(x^-) ~=~ -\frac{1}{a_i(x^-)} \frac{d^2 a_i(x^-)}{d(x^-)^2} \delta_{ij}\,.
\end{align}
In our case the explicit expressions of the mass matrix elements are given by
\begin{align}
\label{eq:cAyphi1}
	\cA_{11} ~=~ - \sqrt{\frac{ G}{r^2\, \Gamma^2\, \Pi_0}}\, \frac{d^2 }{d(x^-)^2}\sqrt{\frac{r^2\, \Gamma^2\, \Pi_0}{ G}}\,,\qquad 
	\cA_{22} ~=~ - \sqrt{\frac{G}{\Pi_0}}\, \frac{d^2 }{d(x^-)^2}\sqrt{\frac{\Pi_0}{G}}\,,
\end{align}
which correspond to the directions along $y$ and $\varphi_1$ respectively, while the remaining six entries are equal and given by%
\footnote{For concreteness we schematically denote with $\cA_{55}$ all $\cA_{kk}$ entries, where  $k = 3,4,\ldots 8$ which includes the $\theta$ and $\varphi_2$ directions. It should be understood that the results presented are valid  for any other  $\cA_{kk}$.}
\begin{align}
	\cA_{55} ~=~ - \frac{1}{\sqrt{\Pi_0}}\, \frac{d^2 }{d(x^-)^2}\sqrt{\Pi_0}\,.
\label{eq:cATor}
\end{align}
These are the elements of the effective mass matrix appearing in the equations of motion \eqref{eq:StringEOM} and thus contain the information about the tidal forces felt by the string as it travels along the null geodesic.
We notice that no direction is flat -- the string feels tidal forces in all spatial directions, including those of the $T^4$. 
Furthermore, since the Penrose limit probes only the immediate neighbourhood of a null geodesic, it is insensitive to the overall size of the manifold.
As a consequence, the mass matrix elements related to the toroidal directions are of the same order as those of the three sphere, despite the warp factor of $T^4$  typically being  parametrically  smaller compared to the warp factors of $S^3$.

\subsection{Tidal effects along the geodesic}

\subsubsection*{Tidal stresses along the toroidal directions}

\begin{figure}[t]
	\centering
	\includegraphics[width=\textwidth]{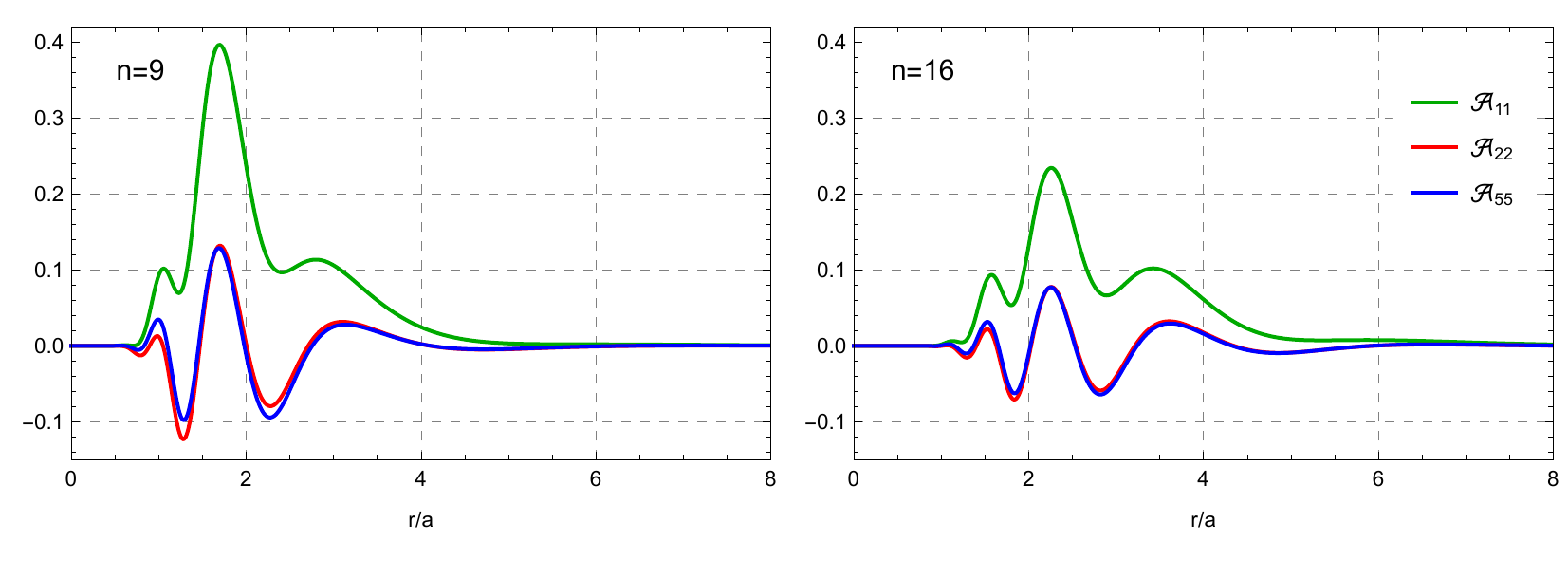}
	\caption{
		Plots of $\cA_{11}$, $\cA_{22}$ and $\cA_{55}$ with  $\tilde b\equiv b/\sqrt{2}a=100$ and $n=9$ (left) and $n=16$ (right).
		The curves are rescaled by $\tilde b^2$ and given in units of $a^2$.
		We observe oscillatory behaviour in all three mass matrix elements. 
		While $\cA_{11}$ is always positive, $\cA_{22}$ and $\cA_{55}$ are negative in some regions.
		The tidal forces along the toroidal directions are not negligible compared to other directions. 
		}
	\label{fig:Aiifig}
\end{figure}

Explicit expressions for $\cA_{ij}$ are not very illuminating thus we do not show them here, however, we present plots of $\cA_{11}$, $\cA_{22}$, and $\cA_{55}$ for some values of the parameters $b$ and $n$ (see  figure~\ref{fig:Aiifig}).
We emphasize that the curves are rescaled by $\tilde b^2\equiv (b/\sqrt{2}a)^2$ and given in units of $a^2$, which is consistent with results in \cite{Tyukov:2017uig}.
Thus when $\tilde b$ is taken to be large in order to approximate the infinite throat of a black hole, the tidal forces become  large as well.
One can observe that the effective mass terms in all directions are of the same order of magnitude. 
Most importantly this implies that \emph{tidal stresses along the toroidal directions felt by the infalling string are not negligible compared to other spatial direction}.

\begin{figure}[t]
	\centering
	\includegraphics[width=\textwidth]{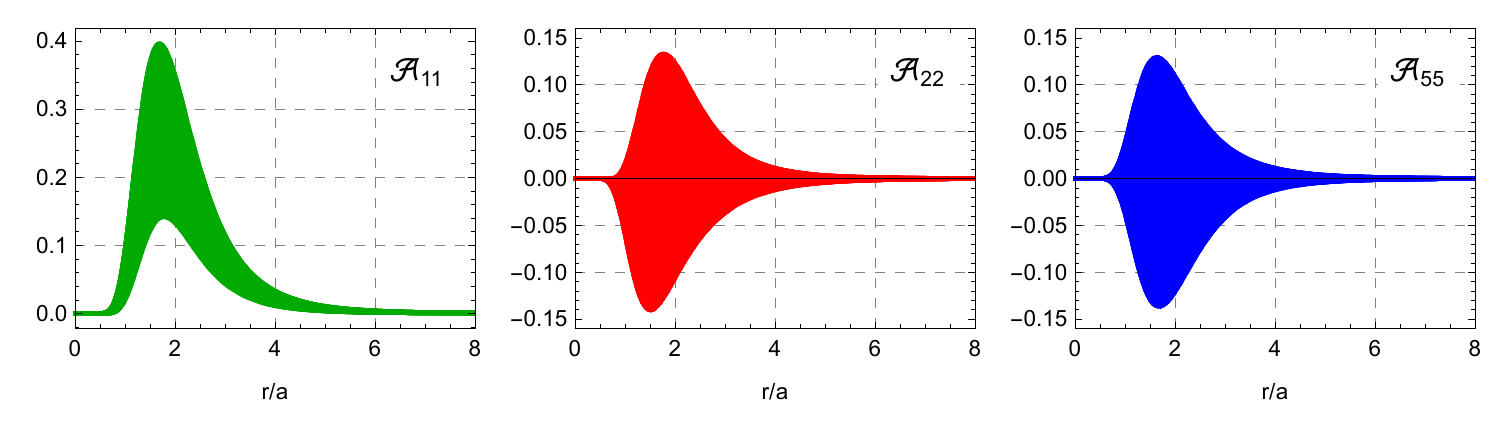}
	\caption{
		Plots of envelopes of $\cA_{11}$ (left), $\cA_{22}$ (middle) and $\cA_{55}$ (right), at  $n=9$. 
		The curves are rescaled by $\tilde b^2\equiv (b/\sqrt{2}a)^2$ and given in units of $a^2$.  In true scale $\cA_{ii}$ become large as $\tilde b$ is increased to obtain a long throat.
	}
	\label{fig:Envelopes}
\end{figure}

We also observe that all mass matrix elements $\cA_{ii}$ are oscillating. 
This is a direct consequence of the spiral motion along the throat and the string frame metric having an  overall conformal factor of $\Pi$ which oscillates as the string moves in the $y$ and $\varphi_1$ directions (see figure~\ref{fig:Pi0Plot}).
The tidal effects $\cA_{11}$ on the string in the $y$-direction are always positive (see also figure~\ref{fig:Envelopes}) -- along this direction the string is always compressed and stabilised.
In the remaining directions  tidal effects oscillate between positive and negative values -- the string feels an alternation of compression and stretching.
When mass matrix elements are negative and large enough massive string modes can get get excited \cite{Martinec:2020cml}. 
In particular, it follows that due to these oscillations, the null geodesic eventually passes through a region  in which $\cA_{55}$ is  negative.
Therefore, according to \eqref{eq:StringEOM}, \emph{it is possible for the infalling string to lose its kinetic energy  by exciting string modes along the $T^4$ directions.}
This  has important consequences in context of the AdS/CFT duality as excitations along the toroidal directions are identified with bosonic fields in the dual CFT. 
We explore this connection further in the next section.

In figure~\ref{fig:Envelopes} we depict the envelopes of the oscillations of the $\cA_{ii}$ functions for $b \gg a$.
The shape of the envelope is determined by the radial direction, $r$, whereas the oscillations come from its motion along $y$ and $\varphi_1$.
As already mentioned, one observes that despite its oscillatory behaviour $\cA_{11}$ is always positive, while $\cA_{22}$ and $\cA_{55}$ can be both positive and negative.
The envelopes, and thus the tidal effects, are extremal around $r\sim\sqrt n\, a$, which is the bottom of the throat where the microstructure of the superstratum is located.

Again, the envelopes in figure~\ref{fig:Envelopes} are rescaled by $\tilde b^2\equiv (b/\sqrt{2}a)^2$ and given in units of $a^2$, meaning that at large $\tilde b$ the amplitudes of  oscillations in $\cA_{ii}$ scale as $\tilde b^2$.
One can explain this by noting that as $\tilde{b}$ grows, the length of the throat increases.  
In  the scaling limit,%
\footnote{More precisely, one has to take $a\rightarrow 0$, while keeping $b$ is fixed.
But rescaling $r$ with $a$ to get to the bottom of the throat, is as if one is sending $b\rightarrow \infty$.
} 
$\tilde{b}\equiv b/\sqrt{2}a\rightarrow\infty$, 
and the throat deepens to infinity while the geometry of the cap remains fixed \cite{Bena:2007qc,Li:2021gbg}.
A particle  dropped from spatial infinity with fixed energy $E$  encounters the mircrostructure at the bottom of the throat with higher kinetic energy and thus its interaction with the tidal forces are greater.

\subsubsection*{The discrete shift-symmetry of the throat length}

\begin{figure}[t]
	\centering
	\includegraphics[width=\textwidth]{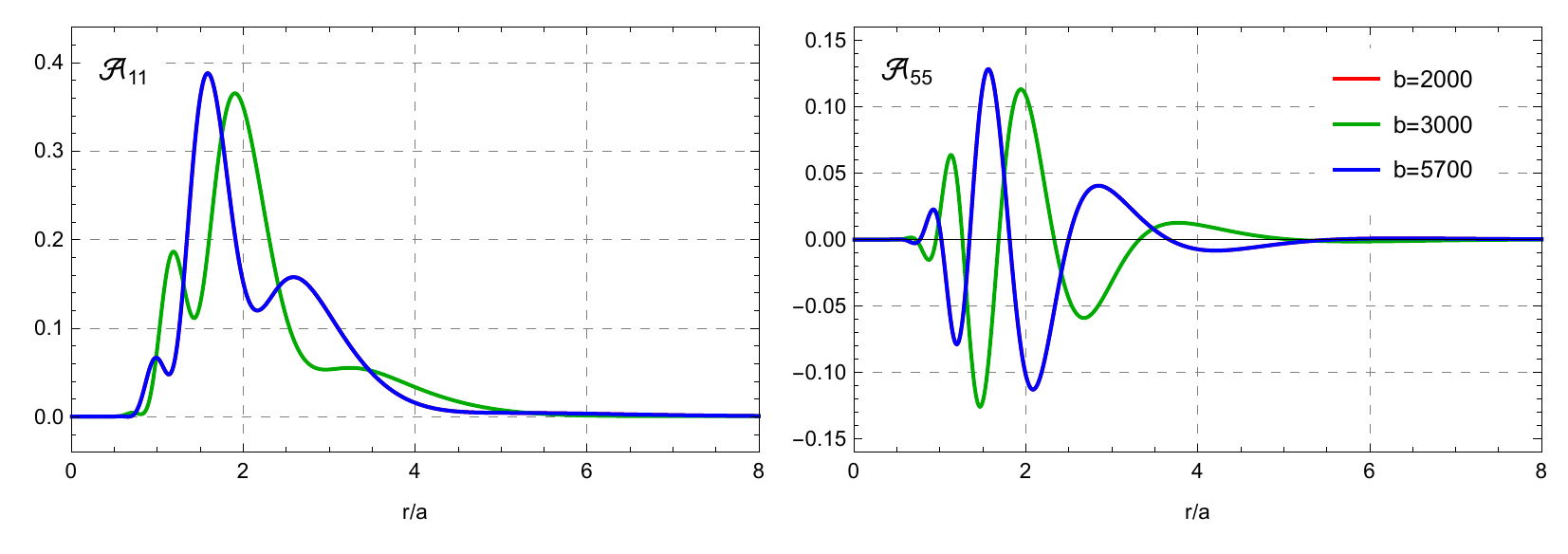}
	\caption{
		Graphs of $\cA_{11}$ and $\cA_{55}$ for  $n=9$ and  $\tilde b = 2000$, 3000 and 5700. 
		The curves are rescaled by $\tilde b^2\equiv (b/\sqrt{2}a)^2$ and given in units of $a^2$.
		For generic values of $\tilde{b}$ the graphs differ, but for specific values of $\tilde b$, such as  $\tilde{b}_0=2000$ and $\tilde{b}_1=5700$, the plots are identical -- in the above plots, the two curves are actually coinciding.}
	\label{fig:bShift}
\end{figure}

As $\tilde b$ increases, the  amplitudes of the tidal forces increase and in addition  the crests of the oscillations within the envelopes shift as well -- see figure~\ref{fig:bShift}.
One notices that after the usual rescaling of the envelope by $\tilde b^2$, the mass matrices  are identical for two  different and carefully tuned throat lengths, $\tilde{b}_0$ and $\tilde{b}_1$. 
In fact, one can show that this pattern continues for an infinite number of throat lengths $\tilde{b}_n$.
The origin of this behaviour is the $v$ and $\varphi_1$ dependent cosine function in $\Pi_0$.
For specific values  of the throat length $\tilde{b}_n$ the argument of that cosine function differs exactly by an integer multiple of $2\pi$ for all values of $\lambda$ (or $r$).
For large $\tilde b$ this means that at specific values $\tilde{b}_n$ the corresponding mass matrix elements $\cA_{ii}$ are identical up to an overall scaling factor of $\tilde b^2$.

As $\tilde{b}\rightarrow\infty$, the family of trajectories of null geodesics in the $\varphi_1$ direction, $\{\varphi_1(\tilde{b};r)\}_{\tilde{b}}$,  converges uniformly  to a continuous function. 
On the contrary, the family of trajectories in the $v$ direction, $\{v(\tilde{b};r)\}_{\tilde{b}}$, converges pointwise to a
limit function which is not integrable and therefore the trajectory $v(r)$ (and thus $y(r)$) varies with $\tilde{b}$.
Consequently, when $\tilde{b}\gg 1$, increasing the throat length   modifies the trajectory in the $v$-direction, while leaving the stabilized trajectory in $\varphi_1$ unchanged.
Furthermore, it follows that the oscillations in all $\cA_{ii}$, induced by the argument in the cosine function in (\ref{eq:BumpDef2}), are mainly due to the motion along the $y$-circle. 
The change in position in $\varphi_1$  does not exceed $\pi$.
This also explains why at large $\tilde b$, the number of oscillations scales with $n$.

Physically, due to a $\mathbb{Z}_{2n}$-symmetry of the superstrata  along $\sqrt2 v$ in the string frame, the specific values of $\tilde b_n$ correspond to  throat lengths such that differences in trajectories along $y$ between two geodesics is an integer multiple of $-\frac{2\pi R_y}{2n}$.
This precisely induces a phase shift that is an integer multiple of $2\pi$ at all $r$ in the argument of the cosine function of $\Pi_0$.
%

\section{CFT perspective on tidal effects}
\label{sec:CFT}

The presence of tidal excitations along $T^4$ can be motivated by considering the dual D1-D5 CFT. It consists of a bound state of D1 and D5 branes wrapping compact dimensions of the theory, namely the $S^1_y$ and $T^4$. More specifically, the D1 branes wrap $S^1_y$ and the D5 branes wrap $S^1_y\times T^4$. The theory is then described by `component' strings composed of the wrapped D-branes which live on the common direction, $S^1_y$, denoted as the $y$ circle. There is a well defined map between the superstrata geometries and states within the D1-D5 system (see \cite{Rawash:2021pik} and references therein). The degrees of freedom correspond to open string excitations which can be either bosonic or fermionic. Bosonic excitations, which we schematically write as $\alpha_{-n}$, are polarized along the torus and fermionic excitations, which we schematically write as $d_{-r}$, are polarized along the torus and  the 3-sphere. These bosonic and fermionic degrees of freedom can be combined into composite operators such as $L_{-n},J_{-n},G_{-r}$ (we have suppressed the charge indices for brevity). The $L_{-n}$'s are Virasoro operators and are written in terms of the energy momentum tensor of the theory. The $J_{-n}$'s are current operators which are written in terms of composite fermionic degrees of freedom. Finally the $G_{-r}$'s are supercharge operators which are written as composites of fermionic and bosonic operators. They exchange fermionic degrees of freedom for bosonic degrees of freedom and vice versa.%
\footnote{For more details, see for example \cite{Avery:2010qw}.}

One can write down the CFT dual of an infalling probe propagating within the superstrata geometry considered in this paper. In order to understand tidal excitations of the probe, which is a dynamical process, we must turn on an interaction in the CFT which is schematically given by the following action
\bea
S~=~ S_0 + \lambda \int d^2w D(w,\bar w)\,,
\eea
where $S_0$ is the action of the free theory. For more work on the deformed CFT see \cite{Avery:2010er,Avery:2010hs,Carson:2014yxa,Carson:2014ena,Carson:2014xwa}. The parameter  $\lambda$ is the coupling of the interaction and $w$ corresponds to the CFT coordinates which describe the location of the operator $D(w,\bar w)$.
This operator consists of two operators which we write schematically as $D =G\sigma$. $G$ is a supercharge operator described previously and $\sigma$ is a twist operator. 
The twist operator joins and splits component strings wrapping the $y$ circle. The twisting-untwisting process applied to component strings in the vacuum generate excitations in the final state. 

To study a probe falling into the superstrata geometry one considers an initial excitation combined with a CFT state which is dual to the superstrata geometry. The CFT dual of the superstrata geometry considered in this paper consists of a left moving momentum wave generated by \cite{Bena:2015bea, Bena:2017xbt}
\bea
|\textrm{background geometry} \rangle ~\sim~ \bigg((L_{-1}-J^3_{-1})^n|00\rangle_1\bigg)^{N_{00}}\bigg(|++\rangle_1\bigg)^{N_{++}}\,.
\eea 
The $|00\rangle_1$ strands correspond to singly wound vacua which contain no net angular momentum along $S^3$ with $N_{00}\propto b^2$. The $|++\rangle_1$ strands correspond to singly wound vacua which contain angular momenta along $S^3$ with $N_{++}\propto a^2$.
The CFT dual of the string probe moving within the superstrata geometry can be schematically given by
\bea
|\textrm{single graviton probe} \rangle ~\sim~ \bigg((L_{-1}-J^3_{-1})^n|00\rangle_1\bigg)^{N_{00}}\alpha_{-n}\bar\alpha_{-n}\bigg(|++\rangle_1\bigg)^{N_{++}}\,,
\label{initial state}
\eea
where the unbarred notation represents a left mover and the barred notation represents a right mover. Note that these excitations correspond to deformations of the metric components in the $T^4$ directions \cite{David:2002wn}. A radial infall with no initial motion along $y$, implies from the CFT perspective, that the left and right movers have the same energy. There is no net momentum along the $y$ circle. 

In order to investigate tidal effects within the CFT, we need a mechanism which takes an initial probe and generates further excitations which grow as a function of time, a process reminiscent of tidal effects experienced by a string in the bulk. In the CFT this effect is produced by acting with a pair of operators, $DD$ on the initial state, (\ref{initial state}) for example. The first operator $D$ twists two smaller component strings in the initial state into a larger component string, and then the second one untwists the larger copy back into two smaller component strings in the final state. By applying two $D$'s, one can compute the transition amplitude for the initial excitation, one left and right moving boson for example, (\ref{initial state}), to split into multiple excitations in the final state. Examples of such amplitudes are given by
\begin{subequations}
	\label{amplitude}
	\bea
	&\langle \textrm{multiple boson state} | DD |\textrm{single graviton probe} \rangle \,,
	\\
	&\langle \textrm{multiple fermion state} | DD |\textrm{single graviton probe} \rangle \,,
	\eea
\end{subequations}
where for example
\bea
|\textrm{multiple boson state} \rangle \,\sim\, \alpha_{-p}\bar\alpha_{-p} \alpha_{-q}\bar\alpha_{-q} \alpha_{-r}\bar\alpha_{-r} |\textrm{background geometry} \rangle
\eea 
corresponds to a final state containing bosonic excitations, and similarly 
\bea
|\textrm{multiple fermion state} \rangle \,\sim\, \alpha_{-p}\bar\alpha_{-p} d_{-q}\bar d_{-q} d_{-r}\bar d_{-r} |\textrm{background geometry}\rangle
\eea
which is a final state containing both bosonic and fermionic excitations. The bosonic mode is required to obtain a non-zero amplitude.
To compute such processes, two $D$ operators are preferred because their combined action ensures that the initial and final twist sectors are the same. This provides the best attempt at maintaining a fixed background to cleanly study processes which happen on top of it. 
Furthermore, we expect such transitions to occur because the twist operator dynamically changes the size of the vacuum. Borrowing intuition from quantum fields in curved space, we know that dynamically changing the vacuum creates particles. The logic here is similar.  In addition, the theory is supersymmetric. Therefore, given an initial state, the operation of $D$ can produce both bosonic and fermionic excitations. 

Recent work \cite{Guo:2021ybz,Guo:2021gqd} suggests that this amplitude should grow in time. This growth indicates the transfer of energy from a single bosonic mode to multiple modes in the final state which share the initial energy amongst themselves. This includes both bosonic and fermionic excitations. The initial mode contains a large momentum in the supergravity picture. This can be pictured as an $L_{-1}$, which functions as a boost generator in AdS, acting many times on an $\alpha_{-1}$ (similarly for the right-moving sector). The multimode final state, however, roughly corresponds in the bulk to a graviton which has lost some of its linear momentum in favour of generating a rest mass, a sign of stringy behaviour. This is suggestive, in the supergravity description, of a string being tidally excited along transverse directions, including the sphere \cite{Martinec:2020cml}. 

Because of the supercharge operator, we expect that transition amplitudes involving the splitting of an initial bosonic excitation into only bosonic excitations in the final state should also grow with time, similar to the amplitude containing fermionic excitations. The bosonic excitations correspond to polarizations along $T^4$. Therefore the CFT process is suggestive of a probe moving within the superstratum geometry and becoming tidally excited along the $T^4$ directions. This is precisely what is suggested by the results of this paper. 
Tidal effects within the CFT are currently being investigated \cite{Guo:2021ybz,Guo:2021gqd}.

\section{Summary and outlook}
\label{sec:CO}

We analysed the the tidal forces along an infalling and spiralling null geodesic in the $\theta = \pi/2$ hypersurface of the $(1,0,n)$ superstata. 
Using the Penrose limit, we found that  there exist tidal forces along all spatial directions orthogonal to the direction of the null geodesic.
This includes tidal forces along the internal $T^4$ directions which 
scale with the length of the throat and are of the same order of magnitude as tidal effects in other, non-internal, directions.
The elements of the mass matrix $\cA_{ii}$ are oscillatory, with the amplitude of oscillation determined by the motion in the radial direction while the oscillatory behaviour coming from motion  along $y$ and $\varphi_1$ directions.

A massless string  travelling along such a null geodesic alternately  experiences  compression and stretching. 
The presence of  stretching regions in the toroidal directions allows for the possibility of string excitations along the $T^4$.
In the dual CFT, such excitations correspond to bosonic degrees of freedom and we discussed the field theory manifestation of tidal forces and their properties.
To conclude, our work supports the hypothesis that the CFT dual of tidal effects on any graviton probe consists in transition amplitudes between different excitations which exhibit a growth behaviour.
The growth in the amplitude for a single-graviton excitation to transition into a final state comprising of multiple bosonic excitations suggests that the initial graviton loses energy, by tidal effects. The bosonic to fermionic amplitudes suggest that the graviton probe loses energy due to tidal excitations along the $S^3$, while the bosonic to bosonic amplitudes concern the energy loss along the $T^4$.

Our analysis suggests a new way in which a string can get trapped inside a capped geometry -- as it spirals down the throat it  encounters tidal forces which transform part of its kinetic energy into internal excitations along the $T^4$ directions. 
The superstratum background can thus be thought of as a viscous fluid in which the string gets captured.

The results presented here are in line with the findings of \cite{Martinec:2020cml}, where tidal excitations were found to excite an infalling string only along the $S^3$ and $S^1_y$ directions. 
This is a consequence of analysing geodesics that lie entirely inside the $\theta = 0$ hypersurface.
The overall conformal factor in the metric, $\Pi$, is then subleading in the Penrose limit ($\Pi\sim 1 + \cO(\Omega^2)$) and thus tidal forces along the $T^4$ are subleading compared to those in the $S^3$ and $S^1_y$ directions.
However, because $\Pi$ is not subleading for any other  infalling  null geodesic, we believe that the takeaway from our analysis applies generically -- tidal forces along $T^4$ are important.
This would suggest that all of the ten dimensions are important when considering the scrambling of strings into capped geometries.

There are still many  questions that need to be answered. 
First, throughout the paper, we have ignored the contribution from the $B$-field. 
This is because in \eqref{eq:StringEOM} it merely mixes different directions, and furthermore,  in the $(1,0,n)$ superstrata the $B$-field has no legs along the $T^4$ directions, hence it should not spoil the results of our analysis.
It is also interesting to observe that the tidal effects are identical between (some) directions in $S^3$ and those of $T^4$, despite possible difference in sizes. 
It would be interesting to see whether subleading effects in the Penrose limit can observe the overall size of individual spatial directions. 
Finally, we have shown that tidal forces along toroidal directions exist and can excite the string, but the exact fate of the string is yet to be determined. 
We hope to answer some of these questions in the future.

\section*{Acknowledgements}

We would like to thank Iosif Bena and Nick Warner for fruitful discussions and useful comments on the draft.
This work is supported by the ERC Grant 787320 - QBH Structure.
YL is partially supported
by the ERC Consolidator Grant 772408-Stringlandscape and the ANR grant Black-dS-String ANR-16-CE31-0004-01.


\bibliographystyle{JHEP}

\bibliography{helix}

\providecommand{\href}[2]{#2}\begingroup\raggedright\begin{thebibliography}{10}

\bibitem{Maldacena:1997re}
J.~M. Maldacena, \emph{{The large N limit of superconformal field theories and
  supergravity}}, {\emph{Adv. Theor. Math. Phys.} {\bfseries 2} (1998) 231}
  [\href{https://arxiv.org/abs/hep-th/9711200}{{\ttfamily hep-th/9711200}}].

\bibitem{Hawking:1974sw}
S.~W. Hawking, \emph{{Particle Creation by Black Holes}},
  \href{https://doi.org/10.1007/BF02345020}{\emph{Commun. Math. Phys.}
  {\bfseries 43} (1975) 199}.

\bibitem{Bena:2016ypk}
I.~Bena, S.~Giusto, E.~J. Martinec, R.~Russo, M.~Shigemori, D.~Turton et~al.,
  \emph{{Smooth horizonless geometries deep inside the black-hole regime}},
  \href{https://doi.org/10.1103/PhysRevLett.117.201601}{\emph{Phys. Rev. Lett.}
  {\bfseries 117} (2016) 201601}
  [\href{https://arxiv.org/abs/1607.03908}{{\ttfamily 1607.03908}}].

\bibitem{Bena:2017xbt}
I.~Bena, S.~Giusto, E.~J. Martinec, R.~Russo, M.~Shigemori, D.~Turton et~al.,
  \emph{{Asymptotically-flat supergravity solutions deep inside the black-hole
  regime}}, \href{https://doi.org/10.1007/JHEP02(2018)014}{\emph{JHEP}
  {\bfseries 02} (2018) 014}
  [\href{https://arxiv.org/abs/1711.10474}{{\ttfamily 1711.10474}}].

\bibitem{Shigemori:2019orj}
M.~Shigemori, \emph{{Counting Superstrata}},
  \href{https://doi.org/10.1007/JHEP10(2019)017}{\emph{JHEP} {\bfseries 10}
  (2019) 017} [\href{https://arxiv.org/abs/1907.03878}{{\ttfamily
  1907.03878}}].

\bibitem{Mayerson:2020acj}
D.~R. Mayerson and M.~Shigemori, \emph{{Counting D1-D5-P microstates in
  supergravity}},
  \href{https://doi.org/10.21468/SciPostPhys.10.1.018}{\emph{SciPost Phys.}
  {\bfseries 10} (2021) 018}
  [\href{https://arxiv.org/abs/2010.04172}{{\ttfamily 2010.04172}}].

\bibitem{Strominger:1996sh}
A.~Strominger and C.~Vafa, \emph{{Microscopic Origin of the Bekenstein-Hawking
  Entropy}}, \href{https://doi.org/10.1016/0370-2693(96)00345-0}{\emph{Phys.
  Lett.} {\bfseries B379} (1996) 99}
  [\href{https://arxiv.org/abs/hep-th/9601029}{{\ttfamily hep-th/9601029}}].

\bibitem{Lunin:2001jy}
O.~Lunin and S.~D. Mathur, \emph{{AdS/CFT duality and the black hole
  information paradox}},
  \href{https://doi.org/10.1016/S0550-3213(01)00620-4}{\emph{Nucl. Phys.}
  {\bfseries B623} (2002) 342}
  [\href{https://arxiv.org/abs/hep-th/0109154}{{\ttfamily hep-th/0109154}}].

\bibitem{Mathur:2005zp}
S.~D. Mathur, \emph{{The fuzzball proposal for black holes: An elementary
  review}}, \href{https://doi.org/10.1002/prop.200410203}{\emph{Fortsch. Phys.}
  {\bfseries 53} (2005) 793}
  [\href{https://arxiv.org/abs/hep-th/0502050}{{\ttfamily hep-th/0502050}}].

\bibitem{Skenderis:2008qn}
K.~Skenderis and M.~Taylor, \emph{{The fuzzball proposal for black holes}},
  \href{https://doi.org/10.1016/j.physrep.2008.08.001}{\emph{Phys. Rept.}
  {\bfseries 467} (2008) 117}
  [\href{https://arxiv.org/abs/0804.0552}{{\ttfamily 0804.0552}}].

\bibitem{Mathur:2009hf}
S.~D. Mathur, \emph{{The information paradox: A pedagogical introduction}},
  \href{https://doi.org/10.1088/0264-9381/26/22/224001}{\emph{Class. Quant.
  Grav.} {\bfseries 26} (2009) 224001}
  [\href{https://arxiv.org/abs/0909.1038}{{\ttfamily 0909.1038}}].

\bibitem{Bena:2018bbd}
I.~Bena, P.~Heidmann and D.~Turton, \emph{{AdS$_{2}$ holography: mind the
  cap}}, \href{https://doi.org/10.1007/JHEP12(2018)028}{\emph{JHEP} {\bfseries
  12} (2018) 028} [\href{https://arxiv.org/abs/1806.02834}{{\ttfamily
  1806.02834}}].

\bibitem{Bena:2015bea}
I.~Bena, S.~Giusto, R.~Russo, M.~Shigemori and N.~P. Warner, \emph{{Habemus
  Superstratum! A constructive proof of the existence of superstrata}},
  \href{https://doi.org/10.1007/JHEP05(2015)110}{\emph{JHEP} {\bfseries 05}
  (2015) 110} [\href{https://arxiv.org/abs/1503.01463}{{\ttfamily
  1503.01463}}].

\bibitem{Ceplak:2018pws}
N.~\v{C}eplak, R.~Russo and M.~Shigemori, \emph{{Supercharging Superstrata}},
  \href{https://doi.org/10.1007/JHEP03(2019)095}{\emph{JHEP} {\bfseries 03}
  (2019) 095} [\href{https://arxiv.org/abs/1812.08761}{{\ttfamily
  1812.08761}}].

\bibitem{Heidmann:2019zws}
P.~Heidmann and N.~P. Warner, \emph{{Superstratum Symbiosis}},
  \href{https://doi.org/10.1007/JHEP09(2019)059}{\emph{JHEP} {\bfseries 09}
  (2019) 059} [\href{https://arxiv.org/abs/1903.07631}{{\ttfamily
  1903.07631}}].

\bibitem{Heidmann:2019xrd}
P.~Heidmann, D.~R. Mayerson, R.~Walker and N.~P. Warner, \emph{{Holomorphic
  Waves of Black Hole Microstructure}},
  \href{https://doi.org/10.1007/JHEP02(2020)192}{\emph{JHEP} {\bfseries 02}
  (2020) 192} [\href{https://arxiv.org/abs/1910.10714}{{\ttfamily
  1910.10714}}].

\bibitem{Shigemori:2020yuo}
M.~Shigemori, \emph{{Superstrata}},
  \href{https://doi.org/10.1007/s10714-020-02698-8}{\emph{Gen. Rel. Grav.}
  {\bfseries 52} (2020) 51} [\href{https://arxiv.org/abs/2002.01592}{{\ttfamily
  2002.01592}}].

\bibitem{Mayerson:2020tcl}
D.~R. Mayerson, R.~A. Walker and N.~P. Warner, \emph{{Microstate Geometries
  from Gauged Supergravity in Three Dimensions}},
  \href{https://doi.org/10.1007/JHEP10(2020)030}{\emph{JHEP} {\bfseries 10}
  (2020) 030} [\href{https://arxiv.org/abs/2004.13031}{{\ttfamily
  2004.13031}}].

\bibitem{Houppe:2020oqp}
A.~Houppe and N.~P. Warner, \emph{{Supersymmetry and Superstrata in Three
  Dimensions}},  \href{https://arxiv.org/abs/2012.07850}{{\ttfamily
  2012.07850}}.

\bibitem{Breckenridge:1996is}
J.~Breckenridge, R.~C. Myers, A.~Peet and C.~Vafa, \emph{{D-branes and spinning
  black holes}},
  \href{https://doi.org/10.1016/S0370-2693(96)01460-8}{\emph{Phys.Lett.}
  {\bfseries B391} (1997) 93}
  [\href{https://arxiv.org/abs/hep-th/9602065}{{\ttfamily hep-th/9602065}}].

\bibitem{Tyukov:2017uig}
A.~Tyukov, R.~Walker and N.~P. Warner, \emph{{Tidal Stresses and Energy Gaps in
  Microstate Geometries}},
  \href{https://doi.org/10.1007/JHEP02(2018)122}{\emph{JHEP} {\bfseries 02}
  (2018) 122} [\href{https://arxiv.org/abs/1710.09006}{{\ttfamily
  1710.09006}}].

\bibitem{Bena:2018mpb}
I.~Bena, E.~J. Martinec, R.~Walker and N.~P. Warner, \emph{{Early Scrambling
  and Capped BTZ Geometries}},
  \href{https://doi.org/10.1007/JHEP04(2019)126}{\emph{JHEP} {\bfseries 04}
  (2019) 126} [\href{https://arxiv.org/abs/1812.05110}{{\ttfamily
  1812.05110}}].

\bibitem{Bena:2020iyw}
I.~Bena, A.~Houppe and N.~P. Warner, \emph{{Delaying the Inevitable: Tidal
  Disruption in Microstate Geometries}},
  \href{https://doi.org/10.1007/JHEP02(2021)103}{\emph{JHEP} {\bfseries 02}
  (2021) 103} [\href{https://arxiv.org/abs/2006.13939}{{\ttfamily
  2006.13939}}].

\bibitem{Bena:2020see}
I.~Bena and D.~R. Mayerson, \emph{{Multipole Ratios: A New Window into Black
  Holes}}, \href{https://doi.org/10.1103/PhysRevLett.125.221602}{\emph{Phys.
  Rev. Lett.} {\bfseries 125} (2020) 22}
  [\href{https://arxiv.org/abs/2006.10750}{{\ttfamily 2006.10750}}].

\bibitem{Bianchi:2020bxa}
M.~Bianchi, D.~Consoli, A.~Grillo, J.~F. Morales, P.~Pani and G.~Raposo,
  \emph{{Distinguishing fuzzballs from black holes through their multipolar
  structure}},
  \href{https://doi.org/10.1103/PhysRevLett.125.221601}{\emph{Phys. Rev. Lett.}
  {\bfseries 125} (2020) 221601}
  [\href{https://arxiv.org/abs/2007.01743}{{\ttfamily 2007.01743}}].

\bibitem{Bena:2020uup}
I.~Bena and D.~R. Mayerson, \emph{{Black Holes Lessons from Multipole Ratios}},
  \href{https://doi.org/10.1007/JHEP03(2021)114}{\emph{JHEP} {\bfseries 03}
  (2021) 114} [\href{https://arxiv.org/abs/2007.09152}{{\ttfamily
  2007.09152}}].

\bibitem{Bianchi:2020miz}
M.~Bianchi, D.~Consoli, A.~Grillo, J.~F. Morales, P.~Pani and G.~Raposo,
  \emph{{The multipolar structure of fuzzballs}},
  \href{https://doi.org/10.1007/JHEP01(2021)003}{\emph{JHEP} {\bfseries 01}
  (2021) 003} [\href{https://arxiv.org/abs/2008.01445}{{\ttfamily
  2008.01445}}].

\bibitem{Bah:2021jno}
I.~Bah, I.~Bena, P.~Heidmann, Y.~Li and D.~R. Mayerson, \emph{{Gravitational
  Footprints of Black Holes and Their Microstate Geometries}},
  \href{https://arxiv.org/abs/2104.10686}{{\ttfamily 2104.10686}}.

\bibitem{Mayerson:2020tpn}
D.~R. Mayerson, \emph{{Fuzzballs and Observations}},
  \href{https://doi.org/10.1007/s10714-020-02769-w}{\emph{Gen. Rel. Grav.}
  {\bfseries 52} (2020) 115}
  [\href{https://arxiv.org/abs/2010.09736}{{\ttfamily 2010.09736}}].

\bibitem{Dimitrov:2020txx}
V.~Dimitrov, T.~Lemmens, D.~R. Mayerson, V.~S. Min and B.~Vercnocke,
  \emph{{Gravitational Waves, Holography, and Black Hole Microstates}},
  \href{https://arxiv.org/abs/2007.01879}{{\ttfamily 2007.01879}}.

\bibitem{Bena:2019azk}
I.~Bena, P.~Heidmann, R.~Monten and N.~P. Warner, \emph{{Thermal Decay without
  Information Loss in Horizonless Microstate Geometries}},
  \href{https://doi.org/10.21468/SciPostPhys.7.5.063}{\emph{SciPost Phys.}
  {\bfseries 7} (2019) 063} [\href{https://arxiv.org/abs/1905.05194}{{\ttfamily
  1905.05194}}].

\bibitem{Bena:2020yii}
I.~Bena, F.~Eperon, P.~Heidmann and N.~P. Warner, \emph{{The Great Escape:
  Tunneling out of Microstate Geometries}},
  \href{https://doi.org/10.1007/JHEP04(2021)112}{\emph{JHEP} {\bfseries 04}
  (2021) 112} [\href{https://arxiv.org/abs/2005.11323}{{\ttfamily
  2005.11323}}].

\bibitem{Martinec:2020cml}
E.~J. Martinec and N.~P. Warner, \emph{{The Harder They Fall, the Bigger They
  Become: Tidal Trapping of Strings by Microstate Geometries}},
  \href{https://doi.org/10.1007/JHEP04(2021)259}{\emph{JHEP} {\bfseries 04}
  (2021) 259} [\href{https://arxiv.org/abs/2009.07847}{{\ttfamily
  2009.07847}}].

\bibitem{Avery:2010qw}
S.~G. Avery, \emph{{Using the D1D5 CFT to Understand Black Holes}},
  \href{https://arxiv.org/abs/1012.0072}{{\ttfamily 1012.0072}}.

\bibitem{Guo:2021ybz}
B.~Guo and S.~Hampton, \emph{{A freely falling graviton in the D1D5 CFT}},
  \href{https://arxiv.org/abs/2107.11883}{{\ttfamily 2107.11883}}.

\bibitem{Guo:2021gqd}
B.~Guo and S.~Hampton, \emph{{The Dual of a Tidal Force in the D1D5 CFT}},
  \href{https://arxiv.org/abs/2108.00068}{{\ttfamily 2108.00068}}.

\bibitem{Blau:2002mw}
M.~Blau, J.~M. Figueroa-O'Farrill and G.~Papadopoulos, \emph{{Penrose limits,
  supergravity and brane dynamics}},
  \href{https://doi.org/10.1088/0264-9381/19/18/310}{\emph{Class. Quant. Grav.}
  {\bfseries 19} (2002) 4753}
  [\href{https://arxiv.org/abs/hep-th/0202111}{{\ttfamily hep-th/0202111}}].

\bibitem{Kanitscheider:2007wq}
I.~Kanitscheider, K.~Skenderis and M.~Taylor, \emph{{Fuzzballs with internal
  excitations}}, {\emph{JHEP} {\bfseries 06} (2007) 056}
  [\href{https://arxiv.org/abs/0704.0690}{{\ttfamily 0704.0690}}].

\bibitem{Bena:2017upb}
I.~Bena, D.~Turton, R.~Walker and N.~P. Warner, \emph{{Integrability and
  Black-Hole Microstate Geometries}},
  \href{https://doi.org/10.1007/JHEP11(2017)021}{\emph{JHEP} {\bfseries 11}
  (2017) 021} [\href{https://arxiv.org/abs/1709.01107}{{\ttfamily
  1709.01107}}].

\bibitem{Bena:2007qc}
I.~Bena, C.-W. Wang and N.~P. Warner, \emph{{Plumbing the Abyss: Black Ring
  Microstates}},
  \href{https://doi.org/10.1088/1126-6708/2008/07/019}{\emph{JHEP} {\bfseries
  07} (2008) 019} [\href{https://arxiv.org/abs/0706.3786}{{\ttfamily
  0706.3786}}].

\bibitem{Li:2021gbg}
Y.~Li, \emph{{Black holes and the swampland: the deep throat revelations}},
  \href{https://doi.org/10.1007/JHEP06(2021)065}{\emph{JHEP} {\bfseries 06}
  (2021) 065} [\href{https://arxiv.org/abs/2102.04480}{{\ttfamily
  2102.04480}}].

\bibitem{Rawash:2021pik}
S.~Rawash and D.~Turton, \emph{{Supercharged AdS${}_3$ Holography}},
  \href{https://arxiv.org/abs/2105.13046}{{\ttfamily 2105.13046}}.

\bibitem{Avery:2010er}
S.~G. Avery, B.~D. Chowdhury and S.~D. Mathur, \emph{{Deforming the D1D5 CFT
  away from the orbifold point}},
  \href{https://doi.org/10.1007/JHEP06(2010)031}{\emph{JHEP} {\bfseries 06}
  (2010) 031} [\href{https://arxiv.org/abs/1002.3132}{{\ttfamily 1002.3132}}].

\bibitem{Avery:2010hs}
S.~G. Avery, B.~D. Chowdhury and S.~D. Mathur, \emph{{Excitations in the
  deformed D1D5 CFT}},
  \href{https://doi.org/10.1007/JHEP06(2010)032}{\emph{JHEP} {\bfseries 06}
  (2010) 032} [\href{https://arxiv.org/abs/1003.2746}{{\ttfamily 1003.2746}}].

\bibitem{Carson:2014yxa}
Z.~Carson, S.~Hampton, S.~D. Mathur and D.~Turton, \emph{{Effect of the twist
  operator in the D1D5 CFT}},
  \href{https://doi.org/10.1007/JHEP08(2014)064}{\emph{JHEP} {\bfseries 1408}
  (2014) 064} [\href{https://arxiv.org/abs/1405.0259}{{\ttfamily 1405.0259}}].

\bibitem{Carson:2014ena}
Z.~Carson, S.~Hampton, S.~D. Mathur and D.~Turton, \emph{{Effect of the
  deformation operator in the D1D5 CFT}},
  \href{https://doi.org/10.1007/JHEP01(2015)071}{\emph{JHEP} {\bfseries 01}
  (2015) 071} [\href{https://arxiv.org/abs/1410.4543}{{\ttfamily 1410.4543}}].

\bibitem{Carson:2014xwa}
Z.~Carson, S.~D. Mathur and D.~Turton, \emph{{Bogoliubov coefficients for the
  twist operator in the D1D5 CFT}},
  \href{https://doi.org/10.1016/j.nuclphysb.2014.10.018}{\emph{Nucl.Phys.}
  {\bfseries B889} (2014) 443}
  [\href{https://arxiv.org/abs/1406.6977}{{\ttfamily 1406.6977}}].

\bibitem{David:2002wn}
J.~R. David, G.~Mandal and S.~R. Wadia, \emph{{Microscopic formulation of black
  holes in string theory}},
  \href{https://doi.org/10.1016/S0370-1573(02)00271-5}{\emph{Phys. Rept.}
  {\bfseries 369} (2002) 549}
  [\href{https://arxiv.org/abs/hep-th/0203048}{{\ttfamily hep-th/0203048}}].

\end{thebibliography}\endgroup

\end{document}